\documentclass[twocolumn,aps,prb,showpacs,superscriptaddress]{revtex4}
\usepackage{graphicx}
\usepackage{color}
\usepackage{mathtools}

\begin{document}
\title{Microscopic examination of hot spots giving rise to nonlinearity in superconducting resonators}

\author{Cihan Kurter}
\affiliation{Center for Nanophysics and Advanced Materials,
Department of Physics, University of Maryland, College Park,
Maryland 20742-4111 USA}

\author{Alexander P. Zhuravel}
\affiliation{B. Verkin Institute for Low Temperature Physics and
Engineering, National Academy of Sciences of Ukraine, 61103 Kharkov,
Ukraine}

\author{Alexey V. Ustinov}
\affiliation{Physikalisches Institut and DFG-Center for Functional
Nanostructures (CFN), Karlsruhe Institute of Technology, D-76128
Karlsruhe, Germany}

\author{Steven M. Anlage}
\affiliation{Center for Nanophysics and Advanced Materials,
Department of Physics, University of Maryland, College Park,
Maryland 20742-4111 USA}

\date{\today}

\begin{abstract}

We investigate the microscopic origins of nonlinear radio frequency
$\it (RF)$ response in superconducting electromagnetic resonators.
Strong non-linearity appearing in the transmission spectra at high
input powers manifests itself through the emergence of jump-like
features near the resonant frequency which evolve towards lower
quality-factor with higher insertion loss as $\it RF$ input power is
 increased. We directly relate these characteristics to the
dynamics of localized normal regions (hot spots) caused by
microscopic features in the superconducting material making up the
resonator. A clear observation of hot spot formation inside a Nb
thin film self-resonant structure is presented by employing the
microwave laser scanning microscope, and a direct link between
microscopic and macroscopic manifestations of nonlinearity is
established.
\end{abstract}


\maketitle
\section{Introduction}
Microwave resonators utilizing superconducting thin films have
enabled development of various nonlinear devices including
bifurcation amplifiers~\cite{Siddiqi}, hot electron
mixers~\cite{Burke}, and kinetic inductance photon
detectors,~\cite{Mazin,Semenov} among other examples. They have been
crucial elements in circuits designed for quantum
computation~\cite{Wallraff}. Recently, these on-chip resonators have
started to appear in metamaterial designs promising new
possibilities such as negative refractive index~\cite{Lazarides},
electromagnetically induced transparency~\cite{Kurter}, tunable THz
response~\cite{Chen}, and parametric amplification of negative index
photons~\cite{Anlage11}.

Nonlinearity is manifested in many forms, including a reduction in
quality factor, an increase in insertion loss, shift of the resonant
frequency~\cite{Anlage99}, or appearance of abrupt jumps in the
transmission/reflection characteristics~\cite{Brenner}.
Understanding the microscopic origin of such a response is
essential. Various mechanisms have been proposed to explain the
nonlinear characteristics at elevated microwave power, including
both intrinsic and extrinsic types. Among them, hot spot formation
is found to be one of the major mechanisms, which was both
theoretically and experimentally studied in superconducting strip
conductors and resonators~\cite{Skocpol}.

A clear and compelling connection between the microscopic models and
the macroscopic picture of nonlinearity revealed by microwave
measurements is still lacking. Here we establish this connection
through a study of the radio frequency $\it (RF)$ response of a
two-dimensional spiral resonator made up of 200 nm Nb thin film on a
quartz substrate via the Laser Scanning Microscopy $\it (LSM)$
technique.
\section{Experiment}
The resonant structures have been designed as engineered {\it
meta-atoms} (planar spirals) of $\it RF$ metamaterials intended to
create a strong magnetic response in the sub-100 MHz
range~\cite{KurterAPL,KurterIEEE}. Prior results show that the
spirals act as very compact self-resonant structures, supporting up
to 10 half-wavelength standing waves of $\it RF$
current~\cite{KurterIEEE}. The {\it free space} stimulation of these
spiral resonators is done by sandwiching them between two $\it RF$
antennas made up of loop-terminated coaxial cables whose other ends
are connected to an $\it RF$ network analyzer. Exciting the spirals
results in Lorentzian type resonant peaks in transmission for low
levels of $\it RF$ stimulus power~\cite{KurterIEEE}. The peaks do
not change in shape or position over a broad range of input powers.
When the spirals are strongly driven, the resonant characteristics
begin to develop notches near the center of the transmission
peaks~\cite{Brenner}. Those nonlinear features evolve with $\it RF$
power and are believed to be associated with the existence of non
superconducting regions in the Nb thin films causing resistive {\it
hot spot} formation.

The high-power properties of superconducting strip resonators are
very sensitive to the geometrical structure of the current-carrying
conductors~\cite{Willemsen}. Our continuous circular spirals have a
unique geometry (see the left inset of Fig.~1) in which the currents
flowing in neighboring turns are parallel, and approximately equal
in magnitude, at least for the first few resonant modes, in contrast
with the current flow profile in meander-line resonators used for
photon counting~\cite{Semenov}. In the spiral, the magnetic fields
generated by flowing currents largely cancel in the region between
the strips, leading to a self-field pattern nearly parallel to the
plane of the spiral. This geometrical effect greatly reduces the
accumulation of $\it RF$ currents at the edges of the strip.
Moreover, the continuous arc of the spiral allows for a relatively
homogenous and uniform flow along the windings compared to
sharp-cornered structures that have a tendency to pile up the
currents at the corners and edges~\cite{Ricci,ZhuravelAPL06}.
\section{Results and Discussion}
\subsection{Global Microwave Transmission Measurements}
The main panel of Fig.~1 shows the global microwave transmission
response of a Nb spiral resonator for a set of $\it RF$ input powers
at 4.4 K. The spiral has an outer diameter, $D_o$ of 6 mm and 40
turns; the strip width, {\it w} and spacing, {\it s} between the
turns are 10 $\mu m$. The fundamental resonance gives a pronounced
transmission peak at $\sim$~75.84 MHz for low and moderate $\it RF$
powers (see Fig.~2 in Ref.~\cite{KurterAPL}). Up to -2 dBm, the
resonant peaks almost overlap, i.e., transmission, $|S_{21}(f)|$ is
a Lorentzian function of frequency and does not show any significant
difference in shape with increasing power (black curves in the right
inset). Those curves falling into the linear response regime
correspond to a state where Nb is in the hotspot-free
superconducting phase. At a critical input power (-1.5 dBm in this
case), $|S_{21}(f)|$ makes a sharp transition from one Lorentzian
curve onto another with higher insertion loss and lower quality
factor, $Q$, as frequency is scanned near resonance (the highest red
curve in the right inset).

As the driving frequency is swept from lower values towards the
resonance frequency, $f_0$, the circulating $\it RF$ power builds up
in the resonator and the circulating currents $I_{circ}$ in the
spiral will increase in magnitude, as $I_{circ}$ =
$I_{max}$/$\sqrt{1+[(f-f_0)/\delta f]^2}$ where $I_{max}$ is the
maximum current induced in the resonator and $\delta$$f$ is the
$3~dB$ bandwidth. For low/medium $\it RF$ input powers, no
instantaneous power value causes excess dissipation because even the
maximum current at resonance is low enough to maintain the linear
characteristics. For the high input power case, the cavity is
over-driven in the vicinity of $f_0$ and thermal effects cause the
breakdown of superconductivity. The over-stimulation of the
resonator launches a new dissipation mechanism and the resonant
response moves onto a different curve with lower $Q$ corresponding
to initial stages of a hot spot. With increasing input power this
transition occurs at progressively lower frequencies where the
dissipative mechanism is activated (the red curves in the right
inset of Fig.~1). At still higher powers, a second discrete
dissipative mechanism is activated, leading to another set of
discontinuous curves (blue curves).  At even more elevated powers
one has either a large number of discrete dissipation centers
activated, or an expansion of the original hot spots, giving rise to
a flattening of the transmission spectrum (orange curves), and
eventually to a dip or crater (brown curves).

The $f_0$ of the Nb spiral remains roughly constant with increasing
power, unlike what is seen in temperature dependence data where the
resonance shifts to smaller frequencies~\cite{Kurter}. At the
extreme case of input power, the dissipation is so large that
superconductivity is quenched and the entire resonant feature is
wiped away from the spectrum (see the data at +27 dBm). Note that
upon reducing the power the resonant features are re-established.

\begin{figure}
\centering
\includegraphics[bb=23 215 596 695,width=3 in]{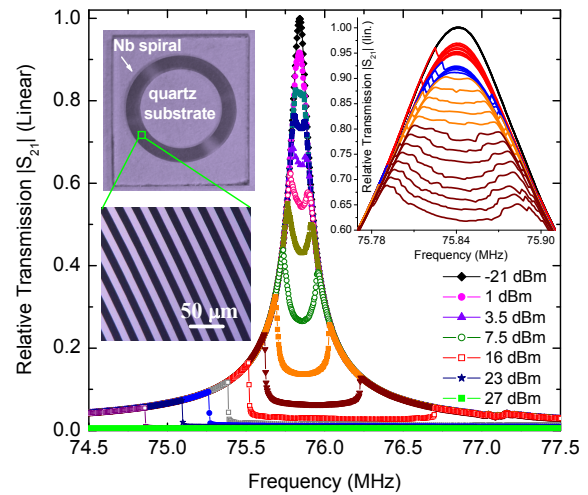}
\caption{(Color online) Global microwave transmission response from
a $D_o$= 6 mm spiral with 40 turns and {\it w} ={\it s} = 10 $\mu m$
for a set of $\it RF$ input powers at $T=$ 4.4 K. Right inset: A
closer look at the linear and nonlinear regimes evolving with
increasing input power. Black Lorentzian type curves corresponding
to power values from -21 dBm to -2 dBm overlap, progressively
evolving into the red curves for power values between -1.5 dBm and
+0.6 dBm then blue curves for powers from +0.7 to +1.1 dBm, orange
curves from +1.2 dBm to +2 dBm and brown curves from +2.2 dBm to
+3.6 dBm. Top Left inset:  An optical image of a spiral resonator
with $D_o$= 6 mm and 40 turns. Bottom Left inset: Zoom in the area
marked in the top left inset showing the Nb spiral windings in
detail. Dark strips are Nb turns in the picture.} \label{Fig1}
\end{figure}
\subsection{Laser Scanning Microscopy}
To develop a microscopic understanding of the nonlinear resonant
features in the transmission data, we examined the spiral resonators
with the $\it LSM$ technique~\cite{ZhuravelAPL06,ZhuravelAPL02}. For
these measurements, the Nb spiral is stimulated by $\it RF$ signals
well below its transition temperature, $T_c$ and is scanned by a
focused laser probe. The spiral resonator is supported on a sapphire
disk along with a thermometer and heater located on a metallic
platform nearby, inside an optical cryostat. The $\it RF$ excitation
has been done by two $\it RF$ loops in a manner identical to that
described above. A diode laser beam of 5 mW power and a wavelength
of 645 nm illuminates the spiral surface. For the large-scale $\it
LSM$ images (up to 10x10 $mm^2$ scanned area with 500x500 points), a
collimated laser beam is focused onto a spot with a diameter of 20
$\mu m$ that produces an estimated temperature difference, $\delta
T$ of 1.2 K at most relative to an area on the sample unexposed to
the laser beam. For more detailed examination of the spiral surface
(within a maximum scanned area of 250 x 250 $\mu m^2$), a 1.6 $\mu
m$ diameter laser beam is used. The raster scanning of the spot is
done with 0.2 $\mu m$ steps, much smaller than the laser spot
diameter. However, the amplitude of the laser is modulated at a
frequency of 100 kHz, which is significantly smaller than the
inverse phonon escape time (on the order of nanoseconds), thus
creating a steady state temperature profile~\cite{ZhuravelAPL02}.
Moreover, the beam intensity is reduced by neutral-density optical
filters down to 1 $mW/mm^2$, so the temperature rise is reduced to
$\delta T$= 0.1 K.

The $\it LSM$ technique basically images photoresponse $(\it PR)$
which comes from the changes induced by periodically modulated laser
power in the transmission characteristics ($\sim$($\partial
|S_{21}(f)|$/$\partial T$)$\delta T$) of the spiral, synchronously
detected with a lock-in amplifier. The response can be decomposed
into inductive and resistive components~\cite{ZhuravelAPL06}. Since
the temperature is well below the $T_c$ of Nb for the measurements
shown here, the modulation of penetration depth $\lambda $ (thus the
inductive response) is small, so the contrast in the $\it LSM$
images is mainly due to resistive $\it PR$, which is proportional to
$\delta(\int R_s \lambda^2 J^2_{RF}(x,y)dS)$, where the integral is
carried out over the area of laser-induced perturbation. The $\it
PR$ is therefore a convolution of changes in resistance below $T_c$,
$\delta R_s$ weighted by the local value of $\it RF$ current density
squared, $J^2_{RF}(x,y)$.

\begin{figure}
\centering
\includegraphics[bb=29 32 598 767,width=3 in]{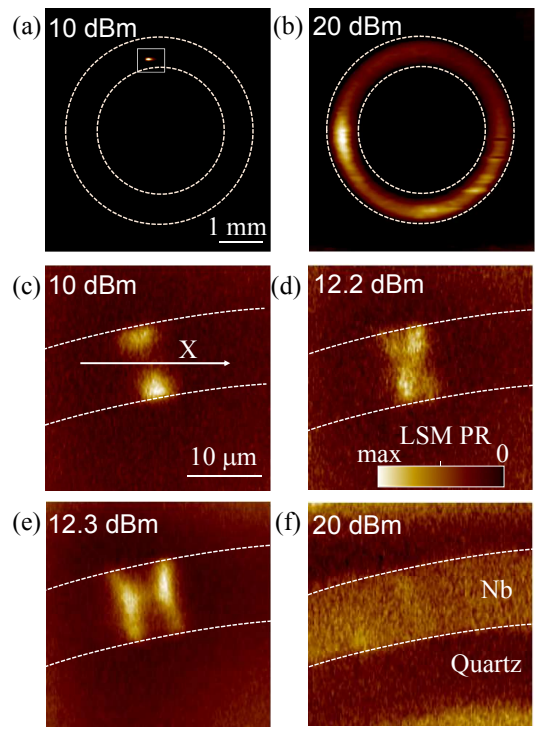}
\caption{(Color online) $\it LSM$ $\it PR$ images of the nucleation
and RF power evolution of a hot spot. (a) Detection of the first
microscopic defect causing the hot spot formation on the Nb spiral
resonator whose details are discussed in Fig.~1 at +10 dBm, 4.8 K.
(b) The hot spot shown in (a) heats the neighboring area and creates
a resistive response coming from the entire spiral at +20 dBm. The
length scale bar in (a) is the same for (b). (c)-(f) Zoom in of the
30x30  $\mu m^2$ area marked in (a) at progressively higher input
powers. The color-scale bar in (d) applies to all LSM images shown
in Fig. 2, the maximum  $\it LSM$ $\it PR$ corresponds to 0.3346 mV
for (a), 0.1857 mV for (b), 3.6 $\mu V$ for (c), 11.4 $\mu V$ for
(d), 23.2 $\mu V$ for (e) and 12.4 $\mu V$ for (f). The length scale
bar in (c) applies to the images in (c)-(f).} \label{Fig2}
\end{figure}

Large-scale (7x7 $mm^2$), low temperature (4.8 K) $\it LSM$ images
of the Nb spiral presented in Fig.~1 are shown in Fig.~2(a) and (b).
The excitation of the spiral is done at the fundamental frequency of
$\sim$ 73.6 MHz~\cite{note1} at which the first emerging hot spot is
detected at +10 dBm (a). By the time the power is ramped up to +20
dBm, the entire sample is heated and the resistive response allows
imaging of a convolution of $\delta R_{s}$ and the current
distribution in the entire spiral (b), showing a single
half-wavelength of standing wave current spanning the length of the
spiral. Most of the $\it PR$ comes from the middle windings, and
diminishes near the inner and outer edges~\cite{KurterIEEE}. The
brighter areas observed on the left side and the lower half of the
spiral in (b) are caused by the structural/morphological defects
with enhanced $\delta R_{s}$.

The upper half of the spiral is free of physical imperfections (no
defect is visible with 0.5 $\mu m$ resolution reflected light
microscope), thus we have concentrated on a small portion of that
area to examine the appearance of the first microscopic defect.
Figs.~2(c)-(f) are a detailed examination of the 30x30 $\mu m^2$
area on the Nb spiral marked in (a) and show the evolution of the
first hot spot with increasing $\it RF$ power at 4.8 K. At +10 dBm,
we observe two bright spots in their initial phase at the opposite
edges of a single strip in the spiral (c), reminiscent of
vortex-antivortex formation at the edges of the current carrying
superconducting stripes~\cite{Huebener}. As the power is increased,
the isolated spots expand and start to merge together (d) where
presumably phase-slip lines (i.e. isothermal nonequilibrium
regions)~\cite{Kuznetsov} are formed. Further increase in $\it RF$
power causes a change of the arrangement of these phase-slip lines
(e) due to the redistribution of current flow along the strip. This
follows from the nucleation of a normal domain across the strip
interfacing with resistive superconducting domains [the bright
regions established along the wings of a butterfly pattern seen in
(e)]~\cite{Sivakov}. With even further increase in power the normal
domain expands along the strip and finally, the entire strip shows a
resistive response (f).

The formation of a phase-slip line in the $\it DC$ resistive state
is signaled by the emergence of step-like features in the $I(V)$
characteristics. Although the phase-slip lines discussed in this
paper are generated by the combination of $\it RF$ excitation and
laser power, we see a very similar picture. The data presented in
Fig.~1 with gradually degrading Q-factor with increasing $\it RF$
input power make a good analogy to the characteristics of the $\it
DC$ bias case with changing differential resistance as the voltage
is increased.

We further  studied the structure of the hot spot with another
sub-100 MHz Nb spiral at 4.5 K. The spiral examined in Fig.~3 has 45
turns, $D_o$ = 5 mm and {\it w} ={\it s} = 10 $\mu m$. Here, the
laser beam passes through the hot spot on a 20 $\mu m$ line scan
[shown with `X' in Fig.~2(c)]. The frequency is swept through
resonance ($f_0$$\sim$ 88.2 MHz) along the X-line cut, to create a
position vs. frequency image of the hot spot at a fixed $\it RF$
input power. Figs.~3(a)-(d) show $\it RF$ power evolution of the hot
spot  within a 20 $\mu m$ line as frequency is scanned in a narrow
band centered on $f_0$, while the corresponding $\it PR$ profiles at
a fixed position, $x_0$ near the center of the hot spot are shown in
(e)-(h) for both directions of frequency sweep. At 0 dBm, in the
initial stages of the hot spot, one sees a strong Lorentzian $\it
PR(f,x_0)$ along with a linear $|S_{21}(f)|^2$ [(a) and (e)],
reminiscent of that at low power shown in Fig.~1. In the case of a
spatially invariant $\delta R_s$, the dissipative term of $\it
PR(f)$ originates from the temperature dependence of $J_{RF}^2(f)
R_s(T)$, meaning that it can be described by a simple difference
between two $|S_{21}(f)|$ characteristics $\delta
|S_{21}(f)|$$\sim$$J_{RF}^2(f)[R_s(T_2)-R_s(T_1)]$, where $T_1$ is
the temperature of the non-radiated area while $T_2$ is the
temperature of the area radiated by the laser probe on the Nb film.
At low $\it RF$ powers, the temperature effect of the laser on
transmission will be small, hence the $\it PR$ traces out a
Lorentzian-like curve, as seen in (e).

As the power increases, the hot-spot produces a nonlinear behavior
of $\delta R_s(J_{RF},T)$ causing peculiar jumps in $|S_{21}(f)|$
whose trend is mimicked by $\it PR(f)$. At +4 dBm [(b) and (f)], in
the middle of the hot spot, one first sees a Lorentzian-like
response at the onset of the frequency sweep, but then a $\it
breakdown$ occurs at a critical frequency, and a $\it crater$
develops in the $\it PR(f)$ trace similar to that seen in the higher
power $|S_{21}(f)|$ characteristics in Fig. 1. Note that these line
cuts are not hysteretic; i. e. forward (blue curves) or backward
(red curves) sweep of the frequency reproduce the same
characteristics.

\begin{figure}
\centering
\includegraphics[bb=14 18 547 763,width=3.4 in]{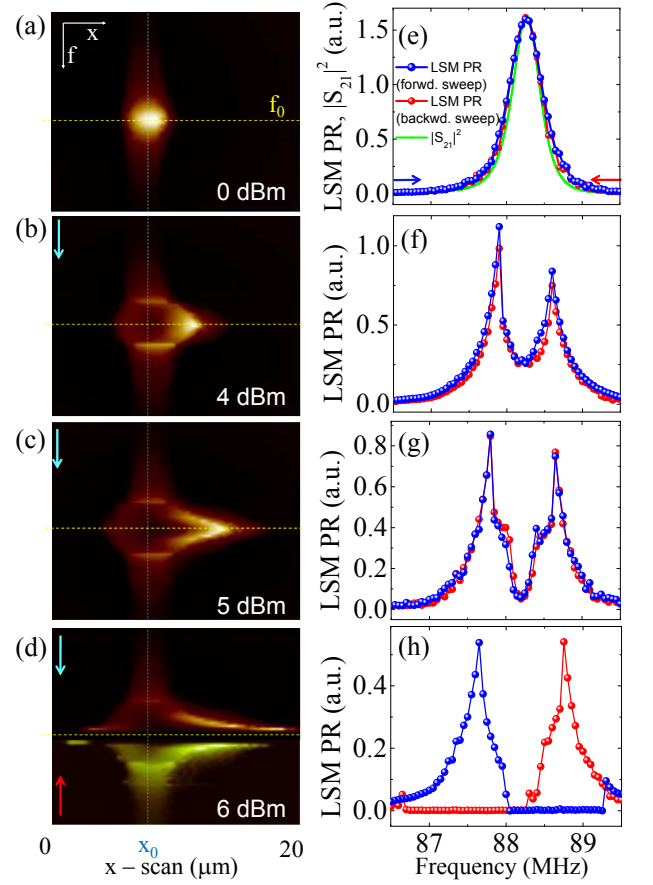}
\caption{(Color online) Frequency and space dependence of $\it LSM$
$\it PR$ at (a) 0 dBm, (b) +4 dBm, (c) +5 dBm (d) +6 dBm on a Nb
spiral with $D_o$= 5 mm, 45 turns, {\it w} ={\it s} = 10 $\mu m$
over a 20 $\mu m$ $x$-scan. The frequency is swept between 86.5 and
89.5 MHz through the fundamental resonant frequency, $f_0$ = 88.2
MHz. The images in (a)-(c) show only forward frequency sweeps whose
direction is indicated with blue arrows. In (d), different shade
color-maps point out both direction sweeps, shown with blue
(forward) and red (backward) arrows. Corresponding $PR(f, x)$
profiles at a fixed position, $x= x_0$ of the laser probe are shown
in (e)-(h). The blue dots are the data collected through forward
frequency sweep whereas the red dots are those through backward
sweep. The $|S_{21}(f)|^2$ at low power, 0 dBm is shown in (e). The
color-scale bar in Fig. 2(d) applies to all LSM images shown in Fig.
3 as well, the maximum $\it LSM$ $\it PR$ corresponds to
(forward/back sweep)-(1.61/1.62) mV for (a); (1.49/1.5) mV for (b);
(1.54/1.55) mV for (c); and (1.12/1.4) mV for (d).} \label{Fig3}
\end{figure}

An important question arises: why does the $\it LSM$ $\it PR$ drop
in the crater?  Is it due to a decrease in $\delta R_s$ or $J_{RF}$,
or both? In fact, both mechanisms should contribute since $\it LSM$
$\it PR$ $\sim$$(\partial R_s$/$\partial T)(\delta T)$+$(\partial
J_{RF}$/$\partial T)(\delta T)$. The dissipative component of $\it
LSM$ $\it PR$ is proportional to both $\delta R_s$ and changes in
the normal (non-superconducting) component of $J_{RF}$, $\delta
J_{RF}$. The latter is reduced in the resistive state due to the
diminished superfluid density. In this case, a crater-like pattern
is formed due to time averaging of $\it LSM$ $\it PR$ coming from
the localized regions where resistive superconducting and normal
domains are side by side, as in phase-slip processes
~\cite{Kuznetsov,Dimitriev}. When a fully normal domain is formed,
$\delta R_s \sim \delta R_N$, where $\delta R_N$ is the very small
change in normal state resistance with temperature, thus the $\it
LSM$ $\it PR$ drops significantly, as seen in (g).

A higher input power (+5 dBm) $\it LSM$ imaging of the same x-line
scan is shown in (c). Note that there are now two shoulders and a
central valley at resonance in the $\it PR$ profile shown in (g).
Our interpretation is that there is a $\it RF$ critical (still
partially superconducting) state with significant $\delta R_s$ for
excitation frequencies on the shoulders. In the $\it PR$ valley, the
material has become normal, mimicking what we have observed in
Fig.~2(e).

With more power, at +6 dBm, the central valley at resonance becomes
deeper (d) and at some point the $\it PR$ drops to zero (h). Though
away from the resonance some recovery is observed, heating is so
significant that back and forth sweeps of frequency can not
reproduce the same $\it PR$ profile. Note the hysteresis in (h); the
direction of the frequency scan is indicated by blue and red arrows.
The size of the resistive domain expands along the x-axis gradually
with increasing power and reaches its maximum at +6 dBm (d).

Line cuts at different $x$ locations in Fig.~3 show a continuous
variation of $\it LSM$ $\it PR(f)$ from Lorentzian-like, to
flat-top, to cratered, which follows what is seen in the global
$|S_{21}(f)|$ measurements shown in Fig.~1. These data provide very
clear observation of such a unique connection between local
temperature change and global $\it RF$-power dependent properties of
a superconducting resonant system.

\begin{figure}
\centering
\includegraphics[bb=50 252 540 709,width=2.3 in]{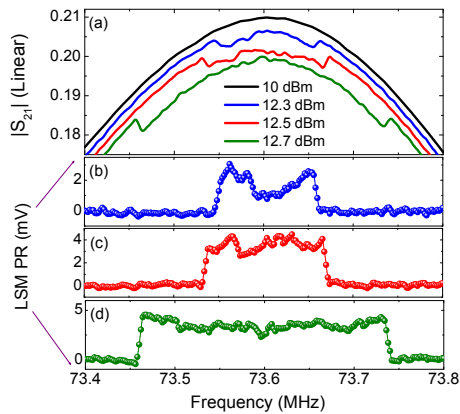}
\caption{(Color online) Evolution of global transmission with small
increments in RF input power (a), Corresponding $\it LSM$ $\it PR$
averaged over 100 $\mu m$ line scans across the center of a $\it
hotspot$. Note that $|S_{21}(f)|$ is measured simultaneously with
$\it LSM$ $\it PR(f)$ when frequency is changed very slowly. A
nearly exact correlation between $|S_{21}(f)|$ and $\it PR$ data is
evident.} \label{Fig4}
\end{figure}

To further support the claims made above, Fig.~4 shows the evolution
of global $|S_{21}(f)|$ characteristics (a) and averaged 100 $\mu m$
line scans of $\it LSM$ $\it PR$ (b)-(d) with increasing power at
4.5 K, on a 40-turn, $D_o$= 6 mm spiral. The linear $|S_{21}(f)|$
characteristics occurring at +10 dBm [black curve in (a)] turns into
nonlinear curves with small increments in $\it RF$ power. The curves
are shifted with equal offset in the vertical direction for clarity.
With increasing input power, the $\it LSM$ $\it PR$ appears over
progressively broader ranges of frequency around resonance
[(b)-(d)]. The development of $\it PR$ contrast at the hot spot
corresponds almost exactly to the deviations of the global
$|S_{21}(f)|$ response from the linear/Lorentzian behavior. This
demonstrates quite clearly the direct correspondence between hot
spot formation and nonlinear response in superconducting resonators.

\section{Conclusions}

In conclusion, we presented experimental evidence linking the global
nonlinear microwave transmission characteristics of a
superconducting resonator with hot spots induced inside the material
making up the resonator. The hot spots are imaged with $\it LSM$
technique at various stages of development and their influence on
the global transmission characteristics is documented.

\begin{center}
ACKNOWLEDGEMENTS
\end{center}

We gratefully acknowledge the contributions of John Abrahams, Tian
Lan, Liza Sarytchev and Brian Straughn. The work at Maryland was
supported by ONR Award No.\ N000140811058 and 20101144225000, the
U.S.\ DOE DESC 0004950, the ONR/UMD AppEl Center, task D10
(N000140911190), and CNAM. The work in Karlsruhe is supported by the
Fundamental Researches State Fund of Ukraine and the German BMBF
under Grant Project No.\ UKR08/011 the DFG-Center for Functional
Nanostructures (CFN), and a NASU program on ``nanostructures,
materials and technologies.''


\begin{thebibliography}{22}
\bibitem{note1} The experimental conditions for the measurements presented in Fig.~1 and Fig.~2 are different, so slight
changes in resonant frequency and critical power are inevitable.
\end{thebibliography}

\end{document}